\documentclass[11pt]{article}
\usepackage{moriond,epsfig}

\newcommand{\etal}{{\it et al.}}

\def\be{\begin{equation}}
\def\ee{\end{equation}}
\def\bea{\begin{eqnarray}}
\def\eea{\end{eqnarray}}

\newcommand{\cf}[1]{{Fig.~\ref{#1}}}
\newcommand{\Br}{{\rm Br}}

\begin{document}
\vspace*{4cm}
\title{Total $J/\psi$ production cross section at the LHC}

\author{J.P. Lansberg }

\address{
Centre de Physique Th\'eorique, \'Ecole Polytechnique, CNRS,   91128 Palaiseau, France
}

\maketitle\abstracts{
We evaluate the production cross section for direct $J/\psi$
integrated in $P_T$ for various collision energies of the LHC in the QCD-based Colour-Singlet Model.
We consider the LO contribution from gluon fusion as well as the one from a fusion
of a gluon and a charm quark from the colliding protons. The rapidity distribution of the yield
is evaluated in the central region relevant for the ATLAS and CMS detectors, as well as in the 
more forward region relevant for the ALICE and LHC-b detectors. The results obtained here 
are compatible with those of other approaches within the range of the theoretical uncertainties
which are admittedly very large. This suggests that the ``mere'' measurements of the yield at the LHC will not help 
disentangle between the different possible quarkonium production mechanisms.
}

\section{Introduction}

In 2007, the first evaluations of QCD corrections to quarkonium-production rates at hadron colliders 
became available. It is now widely accepted -- and understood-- that $\alpha^4_s$ and $\alpha^5_s$ 
corrections to the CSM~\cite{CSM_hadron} are fundamental for understanding the $P_T$ spectrum of 
$J/\psi$ and $\Upsilon$ produced in high-energy hadron  collisions,
\cite{Campbell:2007ws,Artoisenet:2007xi,Gong:2008sn,Artoisenet:2008fc,Artoisenet:2008zza,Lansberg:2008gk} 
while the difficulties of predicting these observables had been initially attributed to 
non-perturbative effects associated with channels in which the heavy
quark and antiquark are produced in a colour-octet 
state~\cite{Lansberg:2006dh,Brambilla:2004wf,Kramer:2001hh,Lansberg:2008zm}.
Further, the effect of QCD corrections is also manifest in the polarisation predictions. While the 
$J/\psi$ and $\Upsilon$ produced inclusively or in association with a 
photon are predicted to be transversally polarised at LO, it has been recently emphasised that their 
polarisation at NLO is  increasingly longitudinal when $P_T$ gets larger. 
\cite{Gong:2008sn,Artoisenet:2008fc,Li:2008ym,Lansberg:2009db,Lansberg:2010vq}

In a recent work,\cite{Brodsky:2009cf} we have also shown that hard subprocesses based on colour 
singlet $Q \bar Q$ configurations alone are sufficient to account for the observed magnitude of 
the $P_T$-integrated cross section. In particular, the predictions at LO~\cite{CSM_hadron} 
(\cf{diagrams} (left)) and NLO~\cite{Campbell:2007ws,Artoisenet:2007xi,Gong:2008sn} 
accuracy are both compatible with the measurements by the PHENIX collaboration at 
RHIC~\cite{Adare:2006kf} within the present uncertainties.\footnote{As recently noted, \cite{Brodsky:2009cf} this points at a reduced impact
of the $s$-channel cut contributions\cite{Haberzettl:2007kj} as well as of the colour-octet mediated channels relevant for the low $P_T$ region. 
The latter are anyway very strongly constrained
by very important recent $e^+e^-$ analyses~\cite{ee} which leave in some cases no room at all for colour octets of any kind.}
 The compatibility between the LO and NLO yields provided  some 
indications that the computations are carried in a proper perturbative regime, at least at RHIC energies.
The agreement with the data is improved when hard subprocesses involving the 
charm-quark distribution of the colliding protons are taken into consideration. These 
constitute part of the LO ($\alpha_S^3$) rate (\cf{diagrams} (right)) and are
responsible for a significant fraction of the observed yield.\cite{Brodsky:2009cf}

We proceed here to the evaluation the $P_T$-integrated yield at higher energies both in the central and forward rapidity regions.
While we find a good agreement with CDF data,\cite{Acosta:2004yw} our study shows that the theoretical uncertainties become very large --close to one 
decade-- reminiscent of the case of total charm production.\cite{Vogt:1900zz} Besides, the yield 
coming from gluon-charm fusion is shown to remain a visible fraction of the direct yield at the LHC
energies. Finally, we shortly discuss the impact of higher QCD corrections and the comparison with other approaches.

\begin{figure}[hbt!]
  \begin{center} 
    \begin{minipage}[b]{0.48\textwidth}%
      \centerline{
 \psfig{figure=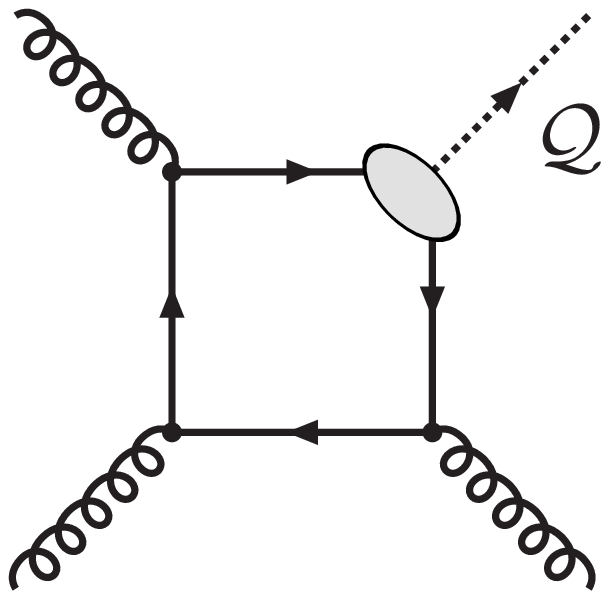,height=3cm}
\psfig{figure=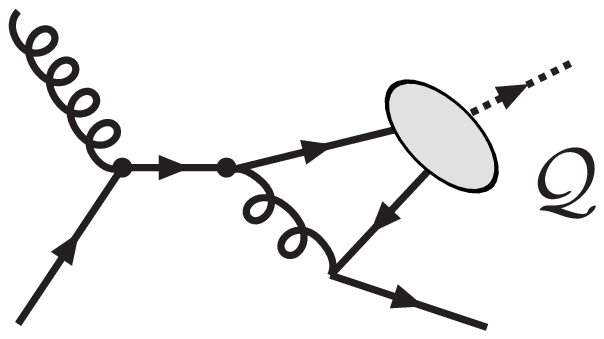,height=3cm}  
      }
      \caption{
Representative diagrams contributing to $^3S_1$ charmonium hadroproduction at high energies in the CSM
 by gluon fusion  (left) and initiated
by a charm quark at order $\alpha_S^3$.
\label{diagrams}
      }
    \end{minipage} 
    \hfill
    \begin{minipage}[b]{0.5\textwidth}%
      \centerline{
   \psfig{figure=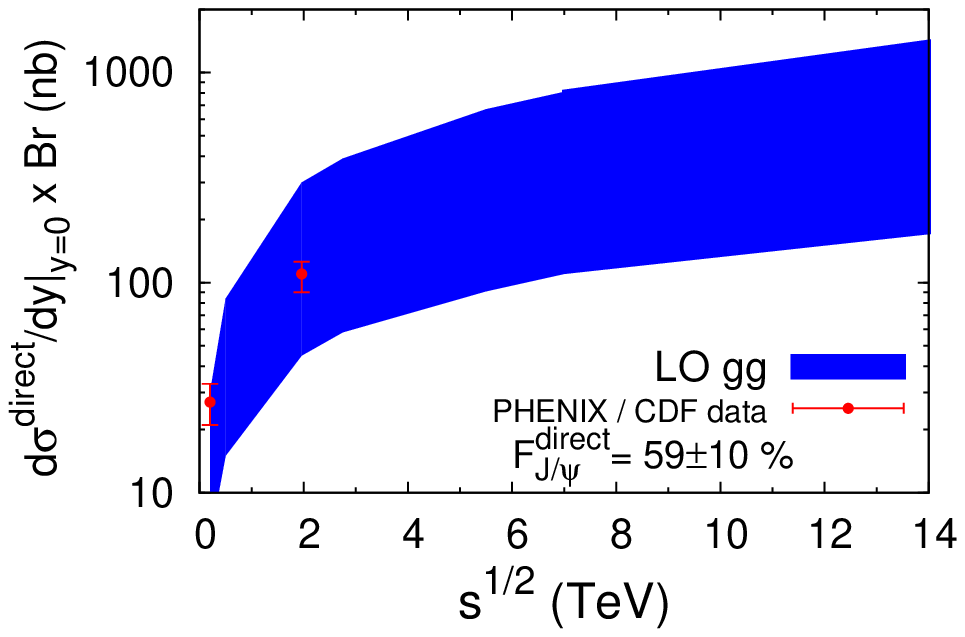,width=\textwidth}  
      }
\caption{$d\sigma^{direct}_{J/\psi}/dy|_{y=0}\times \Br$ from $gg$ fusion in $pp$ collisions for $\sqrt{s}$ from 200 GeV up to 14 GeV
compared to the PHENIX~\protect\cite{Adare:2006kf} and the CDF~\protect\cite{Acosta:2004yw} data  multiplied by the direct fraction.
}
\label{sigma_vs_s}
      \end{minipage}
    \end{center}
\end{figure}

\section{Total $J/\psi$ cross section at the LHC}

The $P_T$ integrated cross sections obtained here have been evaluated along the 
same lines as our previous study.\cite{Brodsky:2009cf} The uncertainty bands have been evaluated 
following exactly the same procedure using the same values for $m_c$, $\mu_R$ and $\mu_F$.

In \cf{sigma_vs_s}, we show $d\sigma^{direct}_{J/\psi}/dy|_{y=0}\times \Br$ from $gg$ contributions as function
of  $\sqrt{s}$ from 200 GeV up to 14 TeV compared to the PHENIX~\protect\cite{Adare:2006kf} and the CDF~\protect\cite{Acosta:2004yw} data multiplied by the direct 
fraction\footnote{Note that the measurement of the prompt yield  by CDF went only down to $P_T=1.25$ GeV. 
We have assumed a fraction of non-prompt $J/\psi$ of $10 \%$ below.}. We have found  a good agreement. At larger energies, these results at 7 TeV (100 to 800 nb) and at 14 TeV (200 to 1400 nb) are
in the same range as those of the Colour Evaporation Model~\cite{Bedjidian:2004gd} with central (upper) values of 140 nb (400 nb) at 7 TeV  and  200 nb (550 nb) at 14 TeV. 
They are also compatible with the results of the ''gluon tower model'' (GTM)~\cite{Khoze:2004eu}, 
300  nb at  7 TeV  and 480 nb at 14 TeV,  
which takes into account some NNLO contributions  shown to be enhanced by $\log(s)$. Quoting the authors,~\cite{Khoze:2004eu}  ``the expected accuracy of the prediction is about a factor of 2-3 in either 
direction or even worse.''

In \cf{7tev}, one shows the differential cross section in rapidity from both $gg$ and $cg$ 
contributions (separately and then summed) at $\sqrt{s}=7$ TeV. One sees that the contribution from $cg$ is not
negligible.  To be more quantitative, we have computed  the ratio $(d\sigma^{cg}_{J/\psi}/dy) / (d\sigma^{cg+gg}_{J/\psi}/dy)$  for $m_c=1.4$ GeV
using 3 choices of the charm distribution in the proton~\cite{Pumplin:2007wg} and taking uncorrelated values for $\mu_R$ and $\mu_F$ for both contributions. At large rapidity, one starts to see 
the enhancement of BHPS~\cite{Brodsky:1980pb} $c(x,Q^2)$ for $x > 0.1$. \cf{LO-2750} and \cf{LO-14000} show the same contributions at $\sqrt{s}=2.75$ TeV and $\sqrt{s}=14$ TeV.

\begin{figure}[hbt!]
  \begin{center} 
    \begin{minipage}[b]{0.48\textwidth}%
      \centerline{
 \psfig{figure=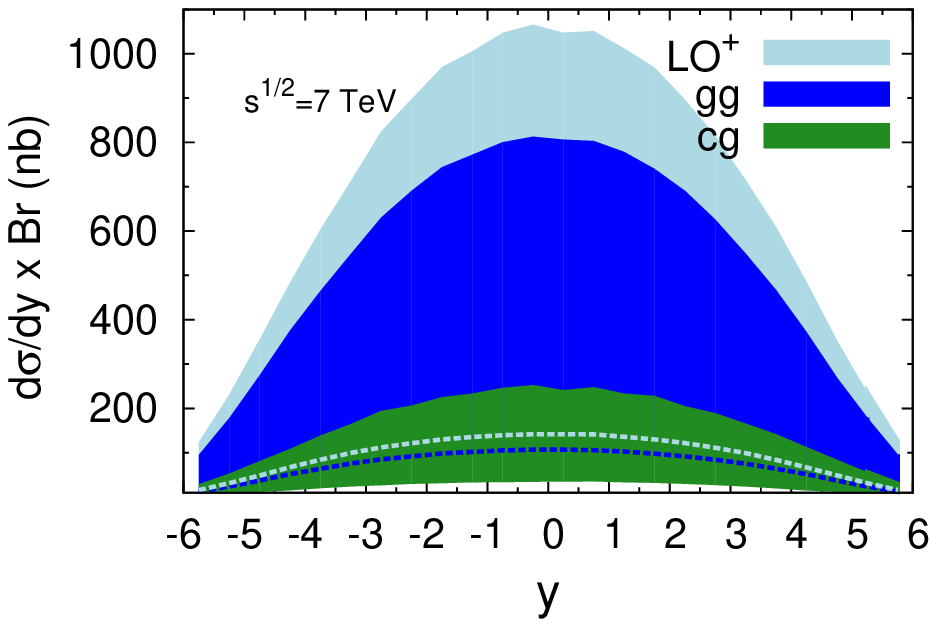,width=\textwidth}
      }
      \caption{$d\sigma^{direct}_{J/\psi}/dy\times \Br$ from $gg$ fusion (dark blue), from $cg$ fusion (green) 
and from all the LO contributions (light blue) in $pp$ collisions at $\sqrt{s}=7$ TeV. \protect\\ 
\label{7tev}
      }
    \end{minipage} 
    \hfill
    \begin{minipage}[b]{0.5\textwidth}%
      \centerline{
   \psfig{figure=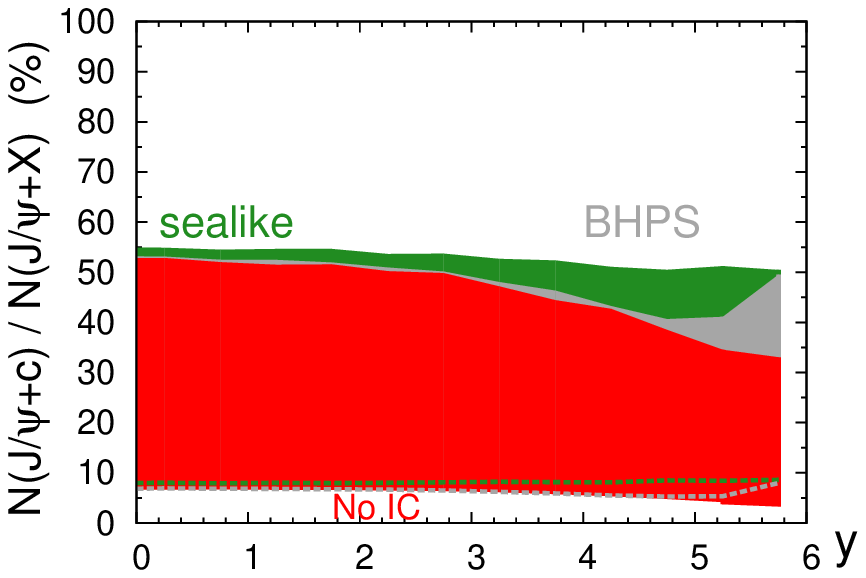,width=\textwidth}  
      }
\caption{Ratio $(d\sigma^{cg}_{J/\psi}/dy) / (d\sigma^{cg+gg}_{J/\psi}/dy)$  at $\sqrt{s}=7$ TeV for $m_c=1.4$ GeV for uncorrelated values of $\mu_R$ and $\mu_F$ for $gg$ and $cg$ contributions and for
3 $c(x,Q^2)$: NoIC (red), sealike (green) and BHPS (gray)}
      \end{minipage}
    \end{center}
\end{figure}

\begin{figure}[hbt!]
  \begin{center} 
    \begin{minipage}[b]{0.48\textwidth}%
      \centerline{
   \psfig{figure=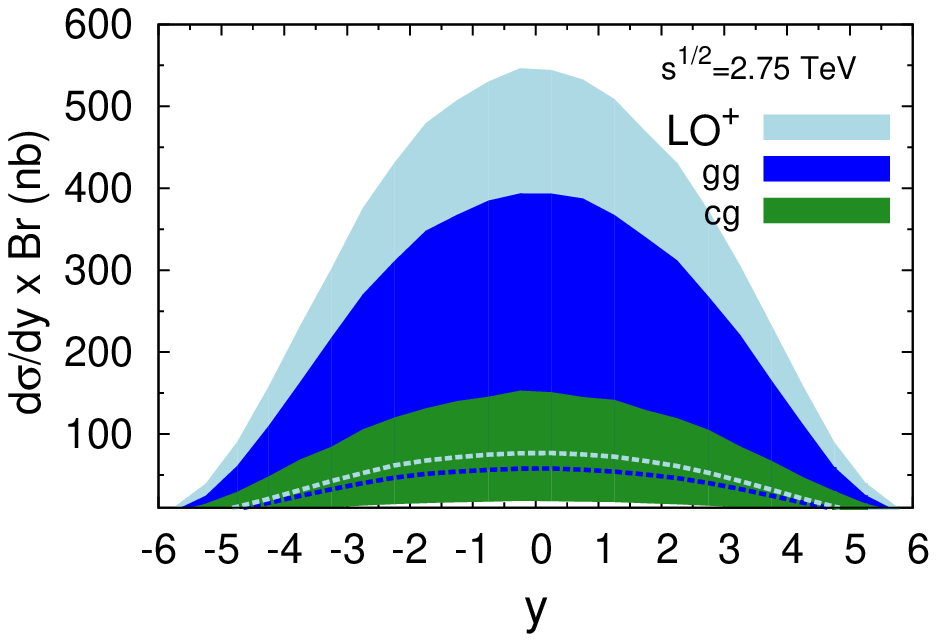,width=\textwidth}  
      }
\caption{$d\sigma^{direct}_{J/\psi}/dy\times \Br$ from $gg$ fusion (dark blue), from $cg$ fusion (green) 
and from all the LO contributions (light blue) in $pp$ collisions at  $\sqrt{s}=2.75$ TeV, i.e. the 
$\sqrt{s_{NN}}$ planned for Pb+Pb collisions in 2010.}\label{LO-2750}
    \end{minipage} 
    \hfill
    \begin{minipage}[b]{0.5\textwidth}%
      \centerline{
 \psfig{figure=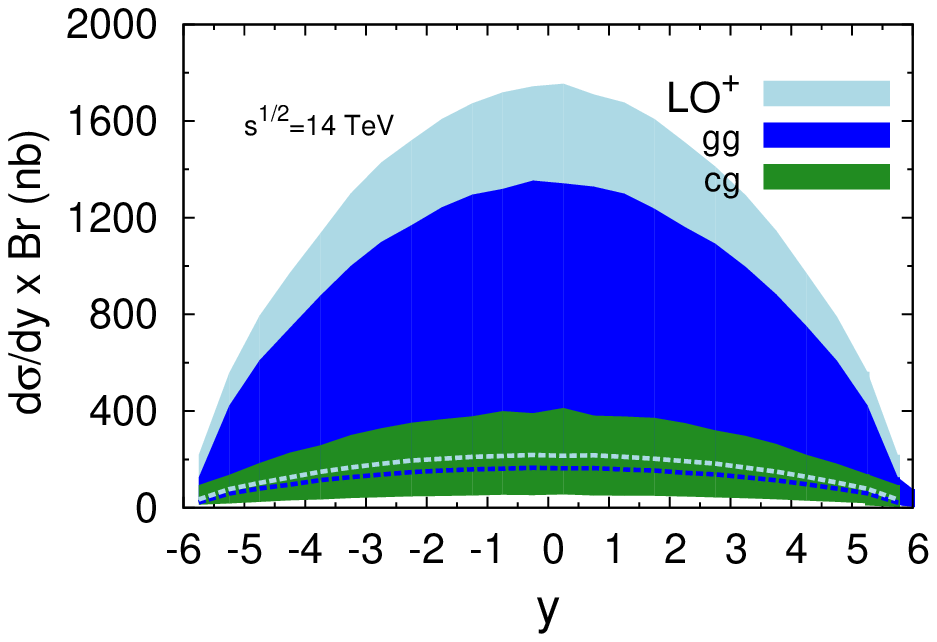,width=\textwidth}
      }
      \caption{$d\sigma^{direct}_{J/\psi}/dy\times \Br$ from $gg$ fusion (dark blue), from $cg$ fusion (green) 
and from all the LO contributions (light blue) in $pp$ collisions at $\sqrt{s}=14$ TeV. \protect\\ 
\label{LO-14000}
      }
      \end{minipage}
    \end{center}
\end{figure}

\section{Discussion and conclusion}

Let us now discuss briefly the expectations for the results when QCD corrections are taken into account. 
First, we would like to stress that, although NLO results~\cite{Campbell:2007ws} are perfectly well behaved 
in nearly all of the phase space region at RHIC energies,~\cite{Brodsky:2009cf} it seems not to be 
so for larger $s$. One observes that the region where the differential cross section in $P_T$ and/or $y$ 
is negative (i.e. very low $P_T$ and large $y$) widens for increasing $s$. Negative differential cross 
section at low $P_T$ is a known issue. Nonetheless, for $\sqrt{s}$ above a couple of TeV, and for some (common) choices of 
$\mu_F$ and $\mu_R$, the $P_T$-integrated ``yield'' happens to become negative, even in the central region. 
This can of course be explained by a larger contribution from the virtual corrections at $\alpha_S^4$ 
--which can be negative-- compared to the real emission contributions --which are positive--. Naturally, 
such results cannot be compared to experimental ones. This also points at likely large virtual NNLO 
contributions at low $P_T$; these are not presently known. Yet, as already mentioned, 
specific NNLO contributions were shown~\cite{Khoze:2004eu} to be enhanced by $\log(s)$.

As we have  discussed above, one may try compare the LO CSM 
with other theoretical approaches such as the CEM~\cite{Bedjidian:2004gd} and the GTM~\cite{Khoze:2004eu}. 
They all qualitatively agree, as well as with PHENIX and CDF measurements. 
For all approaches, one expects a significant spread --up to a factor of ten -- 
of the results when the scales and the mass are varied. 

Owing to these uncertainties, it will be difficult to discriminate between different mechanisms by only 
relying on the yield integrated in $P_T$ and even, to a less extent, on its $P_T$ dependent counterpart. 
This is a clear motivation to study at the LHC other observables related to the production of $J/\psi$ 
such as its production in association with a single charm (or lepton),\cite{Brodsky:2009cf}, with a prompt isolated 
photon~\cite{Li:2008ym,Lansberg:2009db} or even with a pair of $c \bar c$. \cite{Artoisenet:2007xi}

\section*{Acknowledgments}

I would like to thank the organisers for inviting me to their lively conference. I thank V.Khoze, A. Kraan, 
M. Ryskin, G. Smbat, R. Vogt for correspondences and F. Fleuret for useful discussions.

\end{document}